\documentclass[aps,prl,twocolumn,showpacs]{revtex4-1}

\usepackage{color}
\usepackage{graphicx}
\usepackage{amsmath}
\usepackage{amssymb}
\usepackage{bm}
\usepackage{braket}
\usepackage{soul}
\usepackage{gensymb}
\graphicspath{{pict/}{}}

\newcommand\pictc[5]{\begin{figure}
                       \centerline{\vspace{0mm}\includegraphics[width=#1\columnwidth,height=0.7\textheight,keepaspectratio]{#3}}
                       \protect\caption{\protect\label{fig:#4} #5}\vspace{0mm}
                    \end{figure}            }
\newcommand\pict[4][1]{\pictc{#1}{!tb}{#2}{#3}{#4}}
\newcommand\rpict[1]{\ref{fig:#1}}

\begin{document}

\title{Generation of photons with all-optically reconfigurable entanglement \\in integrated nonlinear waveguides}
\author{James~G.~Titchener}
\author{Alexander~S.~Solntsev}
\author{Andrey~A.~Sukhorukov}

\affiliation{Nonlinear Physics Centre, Research School of Physics and Engineering, Australian National University, Canberra ACT 2601, Australia}

\begin{abstract}
We predict that all-optically reconfigurable generation of photon pairs with tailored spatial entanglement can be realized  via spontaneous parametric down-conversion in integrated nonlinear coupled waveguides. The required elements of the output quantum wavefunction are directly mapped from the amplitudes and phases of the classical laser pump inputs in each waveguide.
This is achieved through special nonuniform domain poling, which locally inverts the sign of quadratic nonlinear susceptibility and accordingly shapes the interference of biphoton quantum states generated along the waveguides.
We demonstrate a device configuration for the generation of any linear combination of two-photon Bell states.
\end{abstract}

\pacs{
        42.65.Lm, 	
        42.82.Et,   
		42.50.-p.   
        }

\maketitle

Entanglement is a key characteristic of quantum mechanics, and as shown by Bell  \cite{Bell:1964-195:RAR}, is responsible for quantum theories' violation of the classical principle of local relativistic causality. As well as being of fundamental scientific interest, quantum entanglement can enable powerful new technologies such as quantum computing \cite{Kok:2007-135:RMP}, quantum communication \cite{Gisin:2002-145:RMP}, and quantum enhanced measurement \cite{Giovannetti:2004-1330:SCI}. However the highly advantageous non-classical properties of these technologies will require a flexible interface between the classical and quantum worlds. Indeed the development of such an interface is an area of active research \cite{Yu:2008-233601:PRL, Leng:2011-429:NCOM, Collins:2013-2582:NCOM, Jin:2014-103601:PRL}, the dominant method being spontaneous parametric down-conversion (SPDC), where a classical laser interacts with a  $\chi^{(2)}$ nonlinear medium to produce entangled photons \cite{Yu:2008-233601:PRL, Leng:2011-429:NCOM, Jin:2014-103601:PRL}.

Photons are an ideal platform for creating and manipulating quantum information due to the low noise from the environment and ease of transmission. A qubit encoded into a photon can be easily sent between different photonic elements along an optical fiber, in analogy to the transmission of classical bits along electrical wires \cite{Kok:2007-135:RMP}. Furthermore, logic operations can be preformed on entangled photons by exploiting the nonlinearity inherent in quantum measurement, or by using the interaction between multiple light fields via a nonlinear medium \cite{Kok:2007-135:RMP, Langford:2011-360:NAT}. However in order for these properties to be fully utilized it must first be possible to quickly map information onto the quantum state of entangled photons. Therefore an entangled photon source with fast all-optical reconfigurability of the output wavefunction would be an important step towards realizing new quantum technologies.

\pict{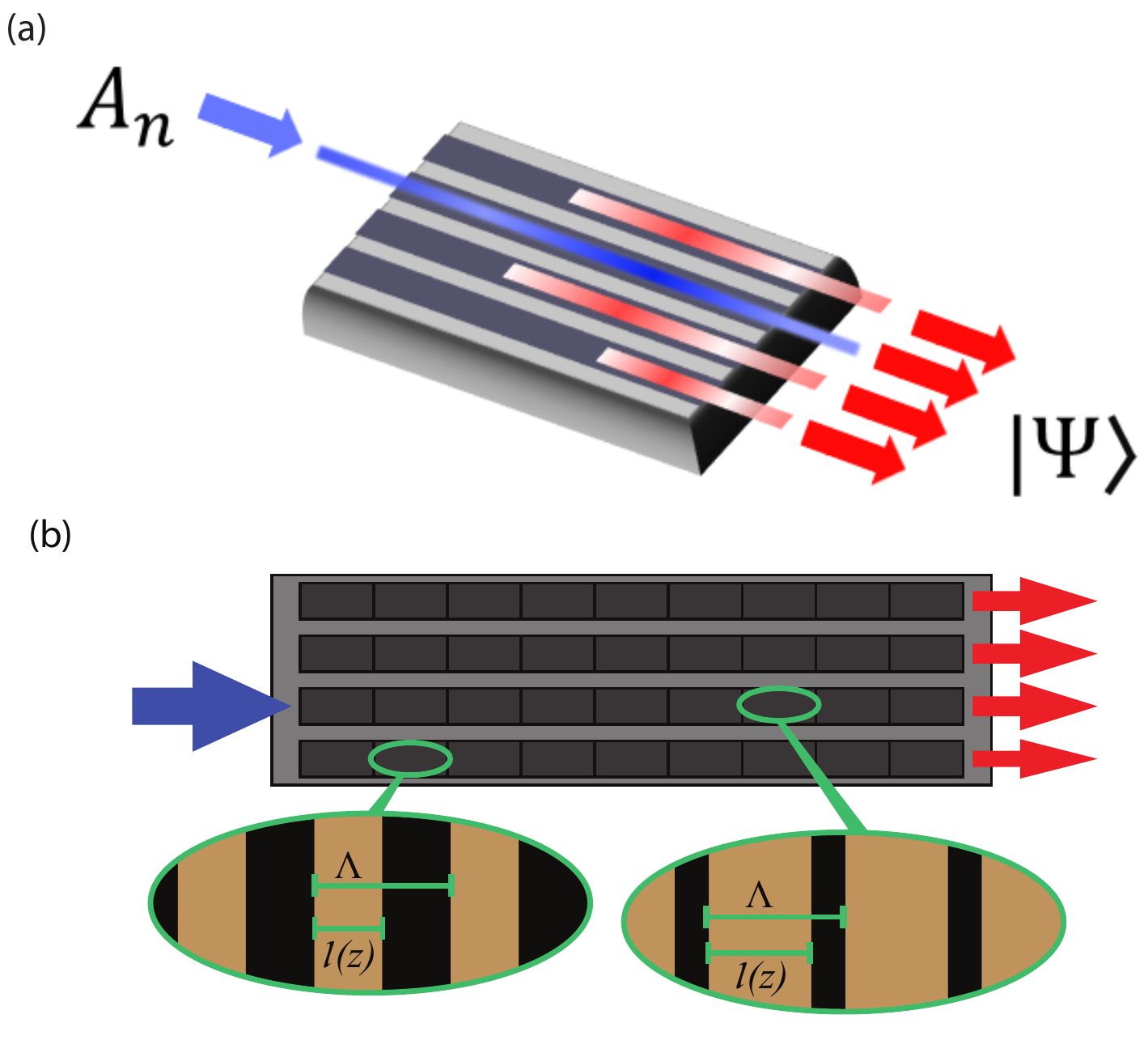}{diagram}{(colour online) (a) Diagram of a nonlinear waveguide array. A pump laser field, $A_{n_{p}}$, shown in blue, drives a waveguide in the nonlinear waveguide array, interaction between the laser and the nonlinearity of the driven waveguide produces entangled photons (red) via SPDC. These entangled photons can couple into neighbouring waveguides, producing path-entangled states, $\Psi$, at the output. (b) Top view of the array. Each waveguide is divided into a number of segments, with different aggregate values of second order nonlinearity. The aggregate nonlinearity is controlled by varying the ratio of up-to-down poling (the duty cycle) of the nonlinear crystal.}

The practicality of any of these technologies will be determined by their scalability and reliability; this suggests the use of quantum photonic circuits will be essential. Integrated circuits can produce and control entangled photons far more efficiently than traditional bulk optics. Integrating optical elements onto a single chip reduces the systems contact with the environment, preserving the fidelity of the quantum entanglement \cite{Politi:2008-646:SCI}. Also integrated devices are compact and stable, so they can be combined to build complex quantum circuits that would be impossibly large using traditional bulk optics. Hence a practical source of entanglement will require the reconfigurable generation and control of entangled photons to be integrated within a photonic chip.

The latest experimental developments \cite{Orieux:2013-160502:PRL, Leng:2011-429:NCOM, Shadbolt:2012-45:NPHOT, Silverstone:2014-104:NPHOT,Jin:2014-103601:PRL} suggest that the goal of a fully integrated and optically reconfigurable source of entangled photons is achievable, but there is still some way to go. For instance \cite{Leng:2011-429:NCOM, Orieux:2013-160502:PRL} show that entangled photons could be created on-chip, while \cite{Shadbolt:2012-45:NPHOT, Silverstone:2014-104:NPHOT} demonstrate that the reconfigurable manipulation of entangled photons in quantum photonic circuits is possible. Indeed the  on-chip generation of photon pairs and their interference using tunable phase shifters was demonstrated in \cite{Silverstone:2014-104:NPHOT, Jin:2014-103601:PRL}, although all-optical reconfigurability as well as the ability to generate a full set of Bell states and their superpositions are yet to be achieved.

In this work we  predict that a photonic chip consisting of an array of coupled nonlinear waveguides \cite{Solntsev:2012-23601:PRL, Solntsev:2014-31007:PRX, Hamilton:2014-83602:PRL} can be designed for all-optically controlled generation of any set of path-entangled  biphoton states. This device is particularly elegant because the output quantum wavefunction is directly mapped from the amplitudes and phases of the classical laser inputs. Hence the device can be reconfigured in real time by varying classical inputs, providing a flexible interface between classical and quantum information.

We illustrate the device concept in Fig.~\rpict{diagram}(a).
Laser driving one of the waveguides in the nonlinear waveguide array (WGA) will generate entangled photon pairs in the driven waveguide via SPDC. The waveguides are coupled such that the biphotons (entangled photon pairs) can tunnel between neighbouring waveguides, but the pumping laser is confined to one waveguide. This is an example of a driven quantum walk \cite{Peruzzo:2010-1500:SCI} of pairs of entangled particles. It has been shown that the interference between the probability amplitudes of different biphoton paths can lead to highly non-classical states at the output of the device \cite{Solntsev:2012-23601:PRL, Solntsev:2014-31007:PRX}. It remained an open question as to whether the WGA could be tuned to produce custom and reconfigurable quantum states, and we demonstrate this can be achieved through specially designed domain poling, as illustrated in Fig.~\rpict{diagram}(b).

We consider a near-degenerate frequency range of the SPDC output, such that the signal and idler photons' frequencies are approximately equal to half the pump frequency, $\omega_{p}$. Then the  state of biphotons within the array is described by the quantum wavefunction
$\Ket{\Psi(z)} = \sum_{n_{s}n_{i}}^{N}\Psi_{n_{s}n_{i}}(z) a^{\dagger}(n_{s})a^{\dagger}(n_{i})\Ket{0}_{s}\Ket{0}_{i}$.
Here $\Psi_{n_{s}n_{i}}(z)$ is the biphoton wavefunction, $z$ is the distance along the propagation direction of the WGA, $a^{\dagger}(n_{s})$ is the creation operator for a signal photon in the waveguide number $n_{s}\in [1,N]$, similarly $a^{\dagger}(n_{i})$ is the creation operator for the correlated idler photon in waveguide $n_{i}$, and $\ket{0}_s\ket{0}_i$ is the vacuum state.

The biphoton wavefunction obeys the differential equation \cite{Solntsev:2014-31007:PRX, Grafe:2012-562:SRP}
\begin{multline}
i\frac{\partial \Psi_{n_{s},n_{i}}(z)}{\partial z} =  i\sum\limits_{n_{p}=1}^{N} A_{n_{p}} d_{n_{p}}(z) e^{i \Delta \beta^{(0)} z} \delta_{n_{s},n_{p}} \delta_{n_{i},n_{p}}  \\-C \bigg[ \Psi_{n_{s},n_{i}+1} + \Psi_{n_{s},n_{i}-1} + \Psi_{n_{s}+1,n_{i}} + \Psi_{n_{s}-1,n_{i}} \bigg].
\label{main}
\end{multline}
 Here the first term on the right is the generation of new biphotons via SPDC, with classical laser driving amplitude $A_{n_{p}}$ and  second order nonlinear coefficient $d_{n_{p}}(z)$ in waveguide number $n_{p}$, while the phase mismatch is denoted as $\Delta\beta^{(0)}$. The last term on the right describes the evanescent coupling of signal and idler photons between neighboring waveguides with the coupling rate given by $C$. Due to the symmetry in Eq.~(\ref{main}), for the initial vacuum state $\Psi_{n_{s},n_{i}}(0)=0$, the biphoton wavefunction is symmetric, $ \Psi_{n_{s},n_{i}}(z)= \Psi_{n_{i},n_{s}}(z)$, at all $z$.

It is important to note that the propagation of biphotons in a nonlinear waveguide array is essentially linear. So if driving the waveguide $n_p$ with unity pump amplitude produces the quantum output state $\ket{\psi_{n_p}}$, then driving multiple waveguides produces the state $\ket{\Psi} = \sum_{n_p=1}^{N} A_{n_{p}}\Ket{\psi_{n_{p}}}$. Therefore by varying the $N$ classical laser inputs, $A_{n_{p}}$, we can reconfigure the device in real time to produce any state in an $N$ dimensional quantum space. However this is only a subspace of the total quantum output space, which will have $N(N+1)/2$ degrees of freedom, namely the probability amplitudes of pairs of photons occupying any two of N waveguides.

We determine that the total output space can be accessed by introducing special domain poling patterns in the WGA [Fig.~\rpict{diagram}(b)]. Adjusting the domain poling pattern allows us to choose the exact form of the $N$ dimensional subspace that is spanned by varying the classical laser inputs to the device. This is achieved by optimizing the poling in each waveguide to produce a specific `basis state', $\ket{\psi_{n_{p}}}$, when it is pumped individually. This method provides an all-optically reconfigurable quantum source.

The typical use of domain poling is to achieve quasi-phase matching (QPM) for $\chi^{(2)}$ nonlinear processes \cite{Fejer:1992-2631:IQE}. QPM involves periodically inverting the orientation of the ferroelectric dipole moment in the nonlinear medium; this corresponds to altering the sign of the second order nonlinearity in the medium.  The technique has also been used to shape the wavefront of down converted photons in bulk nonlinear crystals \cite{Qin:2008-63902:PRL, Leng:2011-429:NCOM, Torres:2004-376:OL, Yu:2008-233601:PRL}. We utilize domain poling within the waveguide array structure to alter the sign of the nonlinear coefficient, $d_{n_{p}}(z)$, as a function of waveguide number, $n_{p}$, and propagation length, $z$. This effectively inverts the quantum phase of biphotons generated at different points along the array, enabling controlled interference between different biphoton paths.

As in QPM we alter the nonlinearity to create poling structure with periodicity that cancels out the phase mismatch of the SPDC process. However we modulate the ratio of up to down poling (the duty cycle) to allow us to tune the efficiency of SPDC, and effectively vary the aggregate value of the nonlinear coefficient along the length of each waveguide. We will show that this 'aggregate' nonlinear coefficient makes it possible to find a general solution for the poling pattern required to produce any arbitrary biphoton wavefunction.
This general solution amounts to solving Eq.~(\ref{main}) for  $\Psi_{n_{s}n_{i}}(L)$ and then inverting the result to express the poling structure, $d_{n_p}(z)$, in terms of $\Psi_{n_{s}n_{i}}(L)$. The role of the aggregate nonlinearity is to mediate between the discrete valued poling structure and the continuous valued biphoton wavefunction.
The solution to Eq. $(\ref{main})$ is found in terms of the discrete Fourier sine transform of the wavefunction, $f_{k_{s},k_{i}}(z)$,

\pict{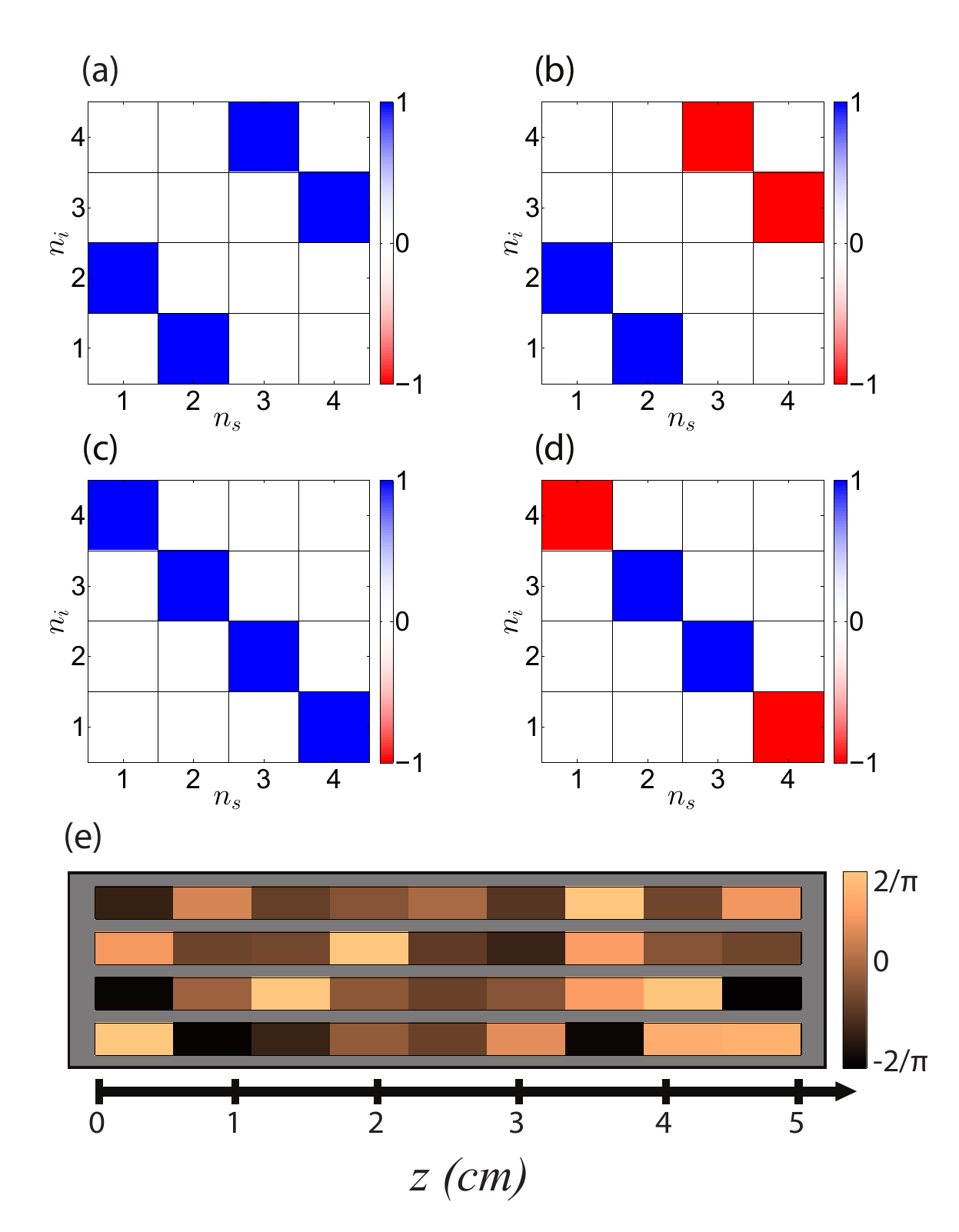}{Fig2}{(color online)
(a),(b),(c),(d)~Target output biphoton states produced when each of the four waveguides is pumped individually. The states are equivalent to Bell states using dual-rail encoding \cite{Shadbolt:2012-45:NPHOT, Kok:2007-135:RMP}, where the signal photon occupying waveguide 1(3) represents a logical 0(1)  and similarly for the idler photon in waveguides 2 and 4.
(e)~The values of the aggregate nonlinear coefficient, $D_{n_{p}}(z)$ along the length of each waveguide. The pump laser interacts with these aggregate nonlinear coefficients to produce the corresponding target outputs in (a),(b),(c),(d). }

\begin{multline}
f_{k_{s},k_{i}}(L)=
 e^{i \beta_{k_{s}k_{i}} L}
 \sum\limits_{n_{p}=1}^N  A_{n_{p}} \sin \left( \frac{\pi k_{i} n_{p}}{N+1} \right) \sin \left( \frac{\pi k_{s} n_{p}}{N+1} \right)\\
 \times  \int\limits_{0}^{L}  d_{n_{p}}(z) e^{i (\Delta \beta^{(0)} - \beta_{k_{s}k_{i}}) z} dz.
 \label{main_sol}
\end{multline}
Here $k_{s}$ and $k_{i}$ are the transverse momenta of the biphotons, $\beta_{k_{s}k_{i}} = 2C \left( \cos(\frac{\pi k_{i}}{N+1})+\cos(\frac{\pi k_{s}}{N+1}) \right) $ is a new phase mismatch term resulting from the transverse momentum of photons within the array, and $L$ is the total length of the array.

In practical implementations of this type of waveguide array the phase mismatch inherent in the waveguide is much larger than the transverse mode phase mismatch, i.e. $ \beta_{k_{s}k_{i}} \ll \Delta \beta^{(0)} $. Therefore in Eq.~(\ref{main_sol}) we can separate the integral over the total phase mismatch into slowly and quickly varying terms, $\beta_{k_{s}k_{i}}$ and $\Delta \beta^{(0)}$ respectively. Under this approximation the integral over $z$ in Eq. $(\ref{main_sol})$ becomes,
  $\int\limits_{0}^{L-\delta L}dz e^{-i\beta_{k_{s}k_{i}}z}\int\limits_{z}^{z+\delta L}d\tau   d_{n_{p}}(\tau) e^{i \Delta \beta^{(0)} \tau}$,
 where $\delta L$ is a length scale  over which the slowly varying term doesn't change significantly but the quickly varying term has one or more complete periods.

We use a QPM poling structure such that $d_{n_{p}}(z)$ is a square wave with periodicity $\Lambda = 2 \pi/\Delta \beta^{(0)}$. This results in the quickly varying term appearing to change linearly over length scales much longer than one period of the poling structure \cite{Fejer:1992-2631:IQE}. The concept of aggregate nonlinearity gives an approximation of the average quantity of biphoton wavefunction generated from a QPM poling structure with arbitrary duty cycle. The aggregate nonlinearity is defined as, $D_{n_{p}}(z) = \Lambda^{-1}\int\limits_{z}^{z+\Lambda} d\tau d_{n_{p}}(\tau) e^{i \Delta \beta^{(0)} \tau} $,
where $\Lambda$ is the quasi-phase matching period.
Over each duty cycle the sign of $d_{n_{p}}(z)$ will change from positive to negative at the point $l_{n_{p}}(z)$. So the aggregate nonlinearity produced by a given duty cycle is

\begin{multline}
D_{n_{p}}(z) = \frac{d_0}{\Lambda}e^{i\phi_{n_{p}}(z) \Delta\beta^{(0)}} \\
\times \left[ \int_{0}^{l_{n_{p}}(z)}e^{i\Delta\beta^{(0)}\tau}d\tau - \int_{l_{n_{p}}(z)}^{\Lambda}e^{i\Delta\beta^{(0)}\tau}d\tau \right].
\label{int}
\end{multline}
Here $d_0$ is the absolute value of the nonlinear coefficient $d_{n_{p}}(z)$, which is unaffected by domain poling. The arbitrary phase $e^{i\phi_{n_{p}}(z) \Delta\beta^{(0)}}$ simply results from translation of each section of the poling structure with respect to the driving laser. Dividing by the poling period, $\Lambda$, ensures that the aggregate nonlinearity represents the amount of biphoton wavefunction produced at an infinitesimal point, rather than the amount produced over a whole period. Integration of (\ref{int}) gives
\begin{multline}
D_{n_{p}}(z) =\frac{2 d_0}{\pi}
\exp\left[     i\Delta\beta^{(0)}      \left( \frac{l_{n_{p}}(z)}{2}+\phi_{n_{p}}(z) \right)    \right]\\ \times\sin\left(\Delta\beta^{(0)}\frac{l_{n_{p}}(z)}{2}\right).
\label{AggNL_sol}
\end{multline}

Now by varying the translation, $\phi_{n_{p}}(z)$, and the length of the positive part of the duty cycle, $l_{n_{p}}(z)$, we can produce an aggregate nonlinearity with any phase and with any magnitude (less than the optimal quasi-phase matching magnitude). Hence by combining a few sections of different duty cycles along the length of each waveguide we can set the aggregate nonlinearity to different values in each section [Fig.~\rpict{Fig2}(e)]. This allows very flexible control over the creation of biphotons and therefore over the final output states of our device.

Substituting the aggregate nonlinearity into Eq. $(\ref{main_sol})$ gives,
\pict{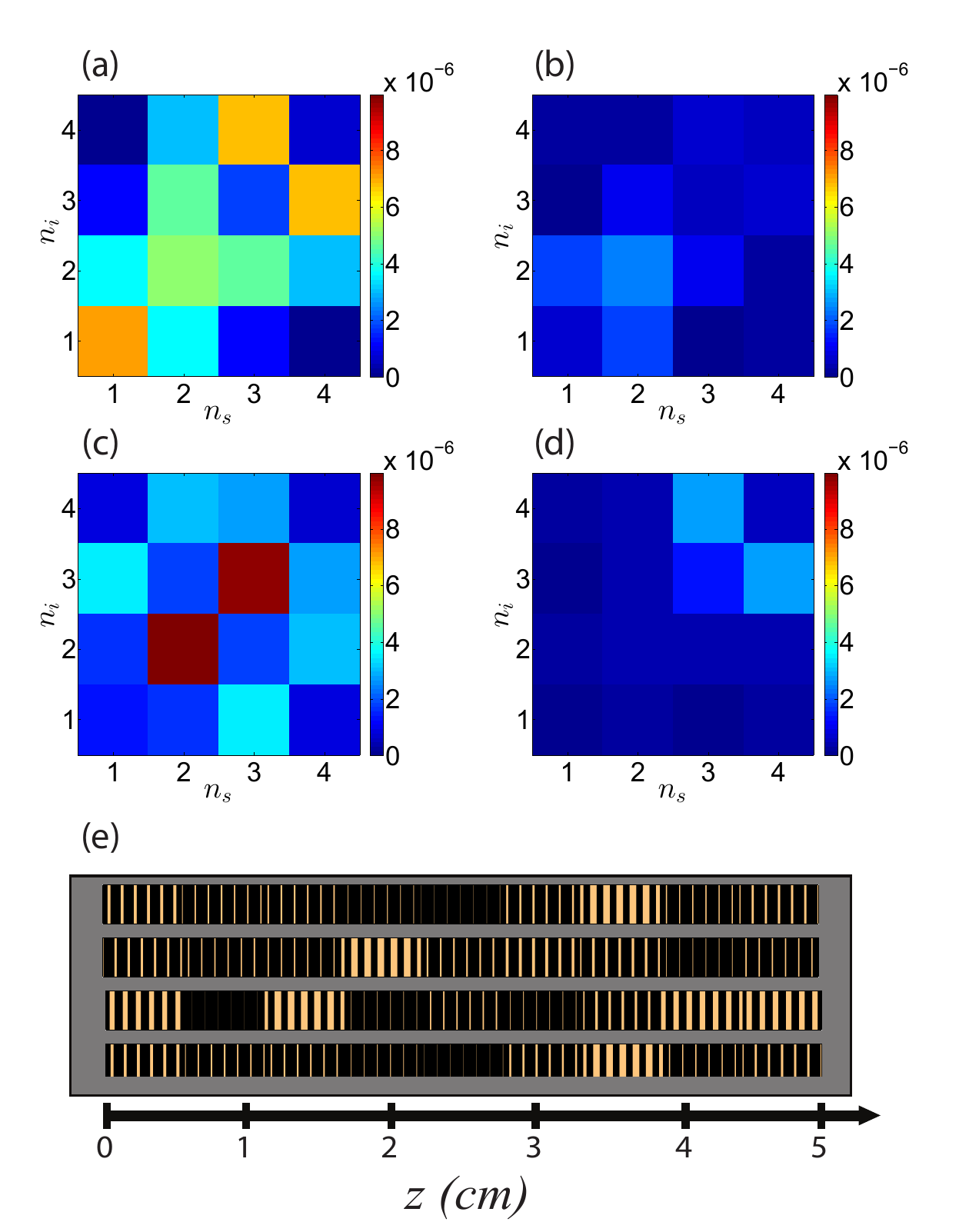}{Fig3}{(color online)
(a),(b),(c),(d)~ The error between the target states in figure \ref{fig:Fig2} (a)-(d) and the states resulting from using the poling structures in figure 3 (e).
(e)~The poling structures used to achieve the aggregate nonlinearities in Fig.~\rpict{Fig2}(e), colours represent the orientation of the ferroelectric dipole moment. Note that the number of domains has been reduced by a factor of around 50 for visualization.
}

\begin{multline}
f_{k_{s},k_{i}}(L)=
 e^{i \beta_{k_{s}k_{i}} L}
 \sum\limits_{n_{p}=1}^N  A_{n_{p}} \sin \left( \frac{\pi k_{i} n_{p}}{N+1} \right) \sin \left( \frac{\pi k_{s} n_{p}}{N+1} \right)  \\
 \times \sum\limits_{j=1}^{M} D_{n_{p}}(z_j) \int_{z_{j}}^{z_{j+1}} e^{-i \beta_{k_{s}k_{i} }z} dz.
 \label{main_sol2}
\end{multline}
This equation can be inverted to solve for the aggregate nonlinearity, $D_{n_{p}}(z_j)$ (see the supplementary material for a detailed discussion of the solvability of the inverse problem). The aggregate nonlinearity is chosen to be an array with number of elements large enough that a solution exists for the target output state $\Psi_{n_{s},n_{i}}(L)$. Generally this will require around $N(N+1)/2$ different values for $D_{n_{p}}(z_j)$ down the length of each waveguide, since this is the dimensionality of the output space.

To illustrate this general approach we design a four-waveguide device with poling to generate the set of two-photon Bell states as the outputs $\ket{\psi_{n_{p}}}$ [Fig.~\rpict{Fig2}(a)-(d)].
Driving the waveguides simultaneously will produce a superposition of the four Bell states, with the amplitude and phase of each Bell state determined by the amplitude and phase of the classical laser driving the waveguide $n_p$.

To be specific, we consider device realized on a lithium niobate (LiNbO$_3$) platform~\cite{Solntsev:2014-31007:PRX}, with
a waveguide length of $L = $5~cm, a coupling rate between the waveguides $C = 161$ m$^{-1}$, and a poling period of $\Lambda =$~18.5~$\mu$m at 230$\degree$C for $775$~nm pump wavelength.
Using these parameters we solve Eq.~(\ref{main_sol2}) for the four Bell states [Fig.~\rpict{Fig2}(a)-(d)],
this provides the required aggregate nonlinearities in the four-waveguide array [Fig.~\rpict{Fig2}(e)].
From the values of the aggregate nonlinearity we can reconstruct the full poling structure, $d_{n_{p}}(z)$, using Eq.~(\ref{AggNL_sol}), the reconstruction is shown in Fig.~\rpict{Fig3}(e).
To check our solutions numerically we compare the states produced by the full poling structure to the target Bell states, the errors are shown in Fig.~\rpict{Fig3}(a)-(d).
The fidelities between the target states and the realized states are all greater than 0.999, this shows that the approximate use of the 'aggregate nonlinear coefficient' is valid for realistic parameters.

Finally we formulate a fast classical characterization  method for the device using stimulated emission tomography \cite{Liscidini:2013-193602:PRL}. By seeding the device with a signal pulse and  measuring the idler output, we can calculate the expected biphoton state produced by SPDC using the transformation

\begin{multline}
\Psi_{n_{i}n_{s}}^{\text{(SPDC)}}=
\sum_{k_{s'},n_{s'}=1}^N \sin \left( \frac{\pi n_{s} k_{s'}}{N+1} \right)
 \sin \left( \frac{\pi n_{s'} k_{s'}}{N+1} \right) \\
  \times e^{(i \beta_{k_{s'}} L)}
 E_{n_{i}n_{s'}}^{(\text{DFG})}.
 \label{final}
\end{multline}
Here $E_{n_{i}n_{s'}}^{(\text{DFG})}$ refers to the idler field produced by difference frequency generation (DFG) in the waveguide number $n_i$ when the seed field is coupled to waveguide $n_s'$. (See supplementary material for a detailed derivation).

In conclusion we have demonstrated how to create all-optically reconfigurable linear combinations of the set of two-photon Bell states in an array of four coupled nonlinear waveguides with special poling. Moreover the poling technique can be applied to a WGA to enable it to produce any set of biphoton states. This opens the door for the design of a variety of reconfigurable entangled photon sources, with output quantum spaces tailored to specific technological applications.
We also note that
the nearest neighbour coupling interactions and nonlinear effects we consider here are common to many physical systems. For instance similar reconfigurable control of entangled photons could be achieved via spontaneous four-wave-mixing in  $\chi^{(3)}$ media.
Further afield, our approach can be extended for controlling the state of pair-correlated atoms generated via spontaneous four-wave-mixing in Bose-Einstein condensates in lattices~\cite{Lewis-Swan:2014-3752:NCOM}.

\pagebreak
\begin{widetext}
\section*{Supplementary Material}
\section{Effect of degeneracies on available output space}

We have shown that a four-waveguide nonlinear waveguide array (WGA) can be used to generate any linear combination of the set of Bell states. In fact domain poling patterns in waveguide arrays provide nearly limitless freedom to produce different quantum states. This, combined with WGA's dynamic all-optical reconfigurability, provides a very flexible source of quantum states of light.

To design a waveguide array capable of being optically reconfigured over a given output space the array must be able to produce the set of 'basis' output states, $\psi_{n_p}$, that define the full output space. If these states can be produced from the device by only varying the pumping lasers, then the device can be reconfigured over the full subspace to any state $\Psi = \sum_{n_p}A_{n_p}\psi_{n_p}$. As we will show in this section, degeneracies in WGA's transverse modes mean that not every subspace defined by a set of arbitary $\psi_{n_p}$ can be spanned. In section \ref{sec:removing} we show that this limitation is removed if the degeneracies are broken by introducing refractive index differences between waveguides.

To investigate the effect of degeneracies on the output space of a WGA it is convenient to consider using domain poling structures in the WGA with different poling periods in different locations. Localized adjustment of the poling period allows phase matching with particular transverse modes of the array. This lets us to isolate and drive each transverse mode in different sections of the array, thus allowing any linear combination of these modes to be produced at the output of the array. Hence any quantum state can be produced at the end of the array given that all the transverse modes can be driven individually. To demonstrate this we consider Eq. (2) from the main publication, showing the output from the array in terms of transverse modes $f_{k_{s},k_{i}}$,

\begin{equation}
f_{k_{s},k_{i}}(L)=
 e^{i \beta_{k_{s}k_{i}} L}
 \sum\limits_{n_{p}=1}^N  A_{n_{p}} \sin \left( \frac{\pi k_{i} n_{p}}{N+1} \right) \sin \left( \frac{\pi k_{s} n_{p}}{N+1} \right)
 \int\limits_{0}^{L}  d_{n_{p}}(z) e^{i (\Delta \beta^{(0)} - \beta_{k_{s}k_{i}}) z} dz.
 \label{main_sol_supp}
\end{equation}

\begin{figure}[t]
\centering
\includegraphics[scale=0.7]{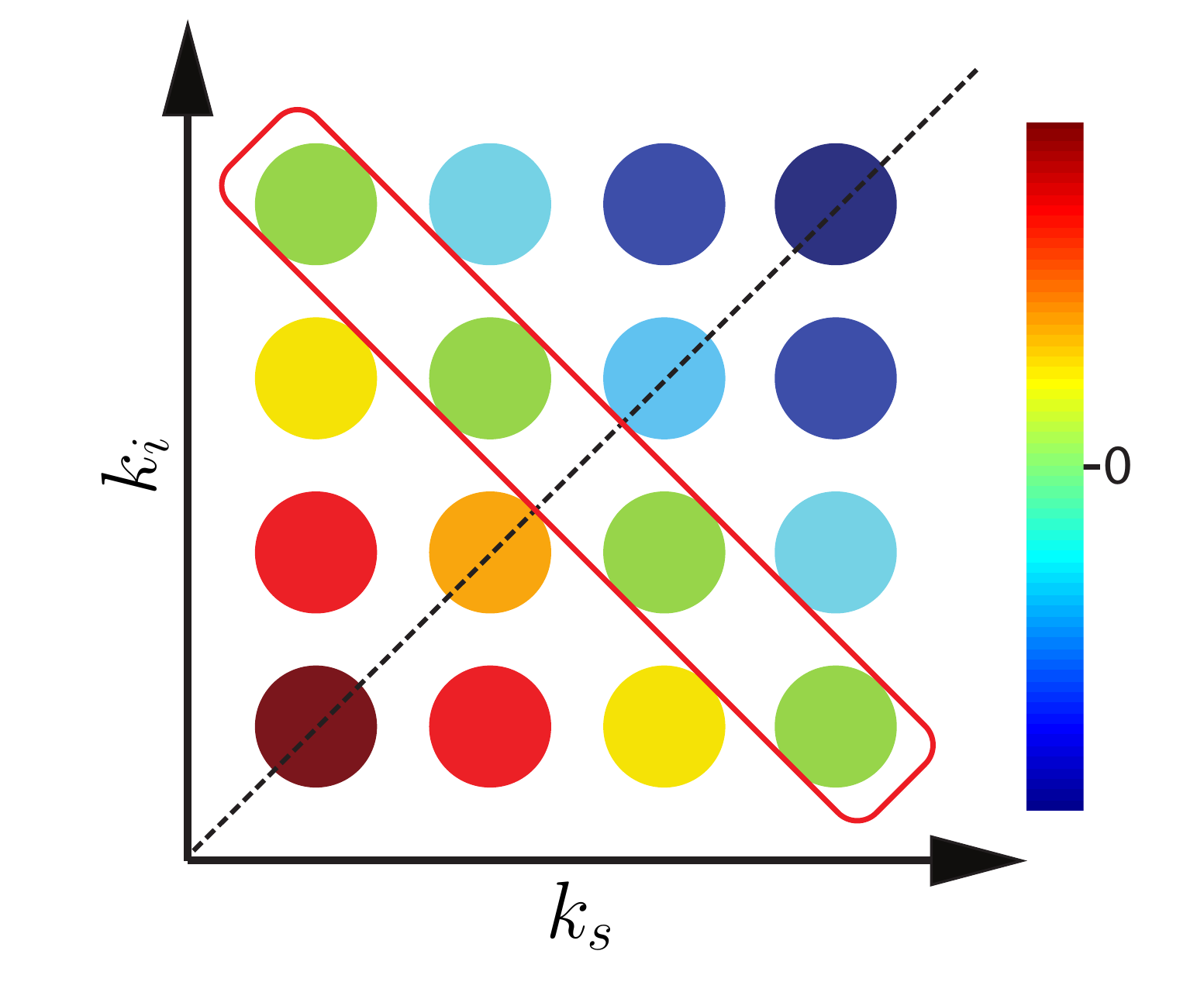}
\caption{Plot of the eigenvalues ($\beta_{k_ik_s}$) of the transverse modes, $f_{k_ik_s}$, for a four-waveguide array. The dashed line shows the axis of symmetry $k_i = k_s$. The red rectangle shows the degenerate eigenvalues along the line $k_s = N+1-k_i$, these all have propagation constants of zero, so cannot be driven individually using varied poling periods.}
\label{fig:first}
\end{figure}

Here we see that poling the nonlinear coefficient, $d_{n_p}(z)$, with a frequency of $\Delta \beta^{(0)} - \beta_{k'_{s}k'_{i}}$ will selectively drive transverse modes that have propagation constants (ie: eigenvalues) $\beta_{k'_{s},k'_{i}}$. Hence transverse modes with unique propagation constants can be phase matched and driven individually, without affecting other modes.
In order to determine which modes have unique propagation constants we must look at the spectrum of the modes eigenvalues as a function of $k_i$ and  $k_s$. In the main text this is given as,



\begin{equation}
\beta_{k_ik_s} = 2 C \left[ \cos \left( \frac{\pi k_i}{N+1} \right) + \cos \left( \frac{\pi k_s}{N+1} \right) \right],
\label{values}
\end{equation}
and a plot of the modes propagation constants for $N=4$ is shown in Fig. \ref{fig:first}. First it is important to note that the equation is transcendental, so there will be some randomly occurring degeneracies, especially for large $N$. There will also be some degeneracies that can be predicted analytically. Here we will consider only the degeneracies that can be predicted analytically, since in WGA's with only a few waveguides these are by far the most common type of degeneracy.

The most obvious of these predictable degeneracies occurs because the signal and idler are indistinguishable, so naturally $\beta_{k_ik_s} = \beta_{k_sk_i}$, and the eigenvalues are symmetric across the main diagonal. This degeneracy is not an obstacle to showing that the device can generate any quantum state because it arises from the particles themselves rather than from the WGA. However there will also be degenerate eigenvalues when $k_s = N+1-k_i$, since in this case $\beta_{k_i,(N+1-k_i)} = 0$ for all $k_i$. This degeneracy corresponds to the set of modes where the entangled signal and idler photons have equal and opposite propagation constants, so the net propagation constant of the state is zero. This presents a problem, because now the degenerate eigenvalues correspond to distinguishable states, but we cannot drive these states independently with domain poling.

As a result of the degeneracy of the $f_{k_i,(N+1-k_i)}$ modes just choosing the domain poling period cannot drive the modes individually. This limitation can be overcome by adjusting the pump intensity in each waveguide. The degenerate transverse modes, $f_{k_i,(N+1-k_i)}$, have distinct spatial profiles for each $k_i$. Hence shaping the spatial profile of the laser driving of the device can drive one of the transverse modes without exciting the others.
Coupling the pump laser into different waveguides will change the rate each of the modes $f_{k_ik_s}$ is driven at. We will call the rate each mode is driven at, $P_{k_ik_s} $, the pump profile. From Eq. (\ref{main_sol_supp}) we can see the pump profile is given by


\begin{equation}
P_{k_ik_s} =\sum_{n_p=1}^{N}A_{n_p} \sin \left( \frac{\pi k_i n_p}{N+1} \right)\sin \left( \frac{\pi k_s n_p}{N+1} \right),
\end{equation}
where $n_p$ denotes the waveguide numbers the pumping lasers are coupled to. Now to control the driving rate of modes with degenerate eigenvalues we should look at $P_{k_i{(N+1-k_i)}}$,

\begin{equation}
P_{k_i{(N+1-k_i)}} = \sum_{n_p=1}^{N}A_{n_p}\sin \left( \frac{\pi k_i n_p}{N+1} \right)\sin \left( \frac{\pi (N+1-k_i) n_p}{N+1} \right).
\end{equation}
This is the rate that each of the degenerate modes $f_{k_i,(N+1-k_i)}$ is driven at when pumping each waveguide with an arbitary intensity and phase pump laser $A_{n_p}$.
It is equivalent to


\begin{equation}
P_{k_i{(N+1-k_i)}} = -\frac{1}{2}\sum_{n_p=1}^{N}A_{n_p}(-1)^{n_p}  \left[ 1 - \cos \left( \frac{2 \pi k_i n_p}{N+1} \right) \right].
\end{equation}

Now if we drive each waveguide with laser field given by $A_{n_p}  = \cos \left( \frac{2\pi k_p n_p}{N+1} \right)(-1)^{n_p} $ then $P_{k_i{(N+1-k_i)}} = \delta_{k_ik_p} + \delta_{(N+1-k_i)k_p}$. Here $k_p$ denotes the mode number of the pump.
This pump profile allows the mode $f_{k'_i{(N+1-k'_i)}}$ to be addressed individually, without exciting the other modes sharing its degenerate propagation constant.
By using a linear combination of these pump profiles, ie
\begin{equation}
A_{n_p}  = \sum_{k_i} c_{k_i} \cos \left( \frac{2\pi k_p n_p}{N+1} \right)(-1)^{n_p},
\end{equation}
 any linear combination of the degenerate modes can be driven. Therefore the degeneracy in the modes propagation constants can be overcome by carefully shaping the pumping, $A_{n_p}$, to exploit the spatial differences between degenerate modes.
 Importantly pumping with the profiles $A_{n_p}  = \sin \left( \frac{2\pi k_p n_p}{N+1} \right)(-1)^{n_p} $ never excites the degenerate $f_{k_i{(N+1-k_i)}}$ modes. So this gives $N/2$ degrees of freedom for exciting other modes while simultaneously using the $\cos \left( \frac{2\pi k_p n_p}{N+1} \right)(-1)^{n_p}$ pumping profile to excite the $N/2$ degenerate $f_{k_i{(N+1-k_i)}}$ modes. This could be useful for allowing reconfigurability, despite degeneracies, by appropriate shaping of the poling structures in the array.

In conclusion we have shown that waveguide arrays can be designed to span large quantum output spaces via all-optical reconfigurability, even in the presence of degeneracies. Through careful choice of which waveguide is driven to produce which basis state, $\psi_{n_p}$, the desired output space can often be spanned. For example in the main publication we show that the set of four Bell states can be spanned with pump excitation of individual waveguides.

\section{Removing degeneracies}
\label{sec:removing}


\begin{figure}[b]
\centering
\includegraphics[scale=0.6]{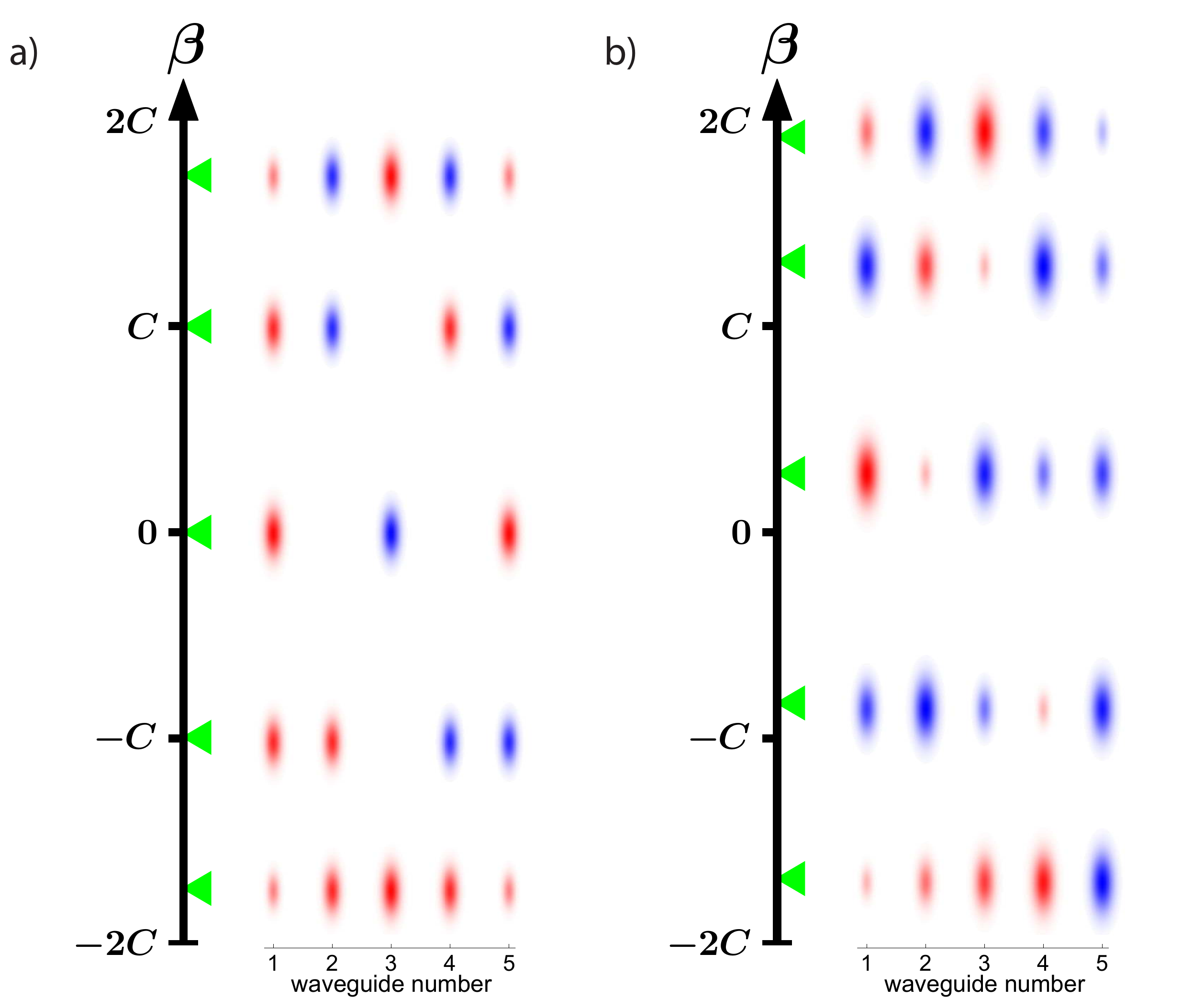}
\caption{ Plots of the single photon modes and their corresponding propagation constants for two different arrays of five waveguides. The array in (a) has uniform refractive index in all waveguides, where as in (b) one of the edge waveguides has a higher refractive index than the others. In (a) the uniform refractive index results in degeneracies in the two photon modes. This can be seen by noting the equal spacing of the single photon modes propagation constants in (a), which, according to Eq. (\ref{cond}) will produce degenerate two photon propagation constants. Also in (a) some transverse modes have no intensity in certain waveguides, meaning they cannot be driven in that waveguide. In (b) we see that both the degeneracies in the propagation constants, and the zero points in the mode profiles have been removed.  }
\label{fig:modes}
\end{figure}

We now consider how degeneracies could be removed from a nonlinear waveguide array to allow it to span any $N$ dimensional biphoton output space. Although, as shown in the main publication, arrays with small numbers of waveguides can span interesting output spaces such as the Bell states, in principle not every output space can be spanned due to degeneracies. This becomes more important for large numbers of waveguides because increasing numbers of modes increases the probability that the transcendental equation for the eigenmodes (Eq. \ref{values}) will produce more degeneracies.

Degeneracies occur when the difference between two single-photon modes eigenvalues are the same. This can be shown by assuming a pair of two-photon modes are degenerate,
\begin{equation}
\beta_{k_{s_1},k_{i_1}}=\beta_{k_{s_2},k_{i_2}} \implies  \beta_{k_{s_1}} + \beta_{k_{i_1}}  = \beta_{k_{s_2}}  + \beta_{k_{i_2}} ,
\end{equation}
then inverting to find a condition relating to the difference between eigenvalues of single-photon modes ,
\begin{equation}
\beta_{k_{s_1}}   - \beta_{k_{s_2}} = \beta_{k_{i_2}}  - \beta_{k_{i_1}} .
\label{cond}
\end{equation}

So to remove degeneracies we require that all the differences between eigenvalues of single-photon modes , $\beta_{k_{s_1}}  - \beta_{s_{s_2}}$, are not equal. This could be achieved by tuning the refractive index in each waveguide to ensure Eq. (\ref{cond}) is violated for all single photon modes. In Fig. (\ref{fig:modes}) we give an example of this degeneracy breaking for an array of five waveguides. This shows that a small adjustment in refractive index can remove degenerate propagation constants as well as removing zero points from transverse modes.

 Allowing for different refractive indices in all waveguides the propagation of biphotons is determined by,

\begin{equation}
i\frac{\partial \Psi_{n_{s},n_{i}}(z)}{\partial z} =  i\sum\limits_{n_{p}=1}^{N} A_{n_{p}} d_{n_{p}}(z) e^{i \Delta \beta^{(0)} z} \delta_{n_{s},n_{p}} \delta_{n_{i},n_{p}}  \\-C \bigg[ \Psi_{n_{s},n_{i}+1} + \Psi_{n_{s},n_{i}-1} + \Psi_{n_{s}+1,n_{i}} + \Psi_{n_{s}-1,n_{i}} \bigg]
+ (\delta\beta_{n_s}+\delta\beta_{n_i})\Psi_{n_s,n_i}
,
\label{RImain}
\end{equation}
where $\delta\beta_{n_s}+\delta\beta_{n_i}$ is the sum of the propagation constants due to refractive index modulation in the waveguides occupied the by signal and idler photons. We now introduce a new set of transverse single-photon modes, $u^{(k)}_{n}$, to account for the new propagation constants in the array. These are eigenmodes of the combined single-photon coupling and propagation operators from Eq. (\ref{RImain}), so for the $k^{th}$ mode

\begin{equation}
\beta_{k} u^{(k)}_{n}  = -C \bigg[ u^{(k)}_{n+1} + u^{(k)}_{n-1}  \bigg]
+ \delta\beta_{n}u^{(k)}_{n}
.
\label{singleRI}
\end{equation}

The two-photon wavefunction produced by down conversion can be expressed in terms of pairs of single-photon modes such that, $\Psi_{n_s,n_i} = \sum_{k_s,k_i}f_{k_s,k_i}u^{(k_s)}_{n_s}\otimes u^{(k_i)}_{n_i}$. Writing Eq (\ref{RImain}) in terms of $f_{k_s,k_i}$ gives,




\begin{equation}
i\frac{\partial f_{k_{s},k_{i}}(z)}{\partial z} =
i D_{k_s,k_i}
+\beta_{k_s,k_i} f_{k_{s},k_{i}}(z),
\label{RIsol1}
\end{equation}
Here $\beta_{k_s,k_i} =(\beta_{k_s} + \beta_{k_i})$ and $D_{k_s,k_i}$ is the spatial overlap between the pumping and the real space profile of the two-photon mode,
 $D_{k_s,k_i} =
 \sum\limits_{n_{p}n_{s}n_{i}}^{N}  A_{n_{p}} u^{(k_s)}_{n_s}\otimes u^{(k_i)}_{n_i} \delta_{n_{i},n_{p}} \delta_{n_{s},n_{p}} d_{n_{p}}(z) e^{i \Delta \beta^{(0)} z}$.
  The solution to Eq. (\ref{RIsol1}) is
\begin{equation}
f_{k_{s},k_{i}}(L) =
e^{i\beta_{k_s,k_i}L}
 \sum\limits_{n_{p}}^{N}
  A_{n_{p}}
  u^{(k_s)}_{n_p}\otimes u^{(k_i)}_{n_p}
 \int_0^L d_{n_{p}}(z)  e^{i \left( \Delta \beta^{(0)} -\beta_{k_s,k_i} \right)z}.
\label{sol}
\end{equation}
This is analogous to Eq (\ref{main_sol_supp}), but now with a new set of modes and eigenvalues.
From this equation the two conditions to exclusively drive any two-photon mode in any waveguide are evident.

\begin{enumerate}
  \item The mode must have a unique eigenvalue, $\beta_{k_s,k_i}$, so that mode can be selectively driven by poling at the resonant spatial frequency $\Delta\beta^{(0)}-\beta_{k_s,k_i}$.
  \item It must be possible to drive the mode by pumping any waveguide, so the overlap of the mode, $D_{k_s,k_i}$, must be nonzero, when each waveguide is driven individually.
\end{enumerate}

The first condition ensures that each mode can be driven exclusively, while the second condition ensures that this exclusive driving can occur in any waveguide. In Fig. (\ref{fig:modes}) it is shown that both criteria are satisfied for an array of five waveguides, but only after one waveguide is given a different refractive index to the others, so that $\delta\beta_n = 1$ for one waveguide in Eq. (\ref{RImain}). These two criteria make it possible to design a WGA to span any desired $N$ dimensional output space.

This would be achieved by poling each waveguide with $N(N+1)/2$ different segments of poling. Each segment being poled with a period $2\pi/(\Delta\beta^{(0)}+\beta_{k_s,k_i})$ to phase match a particular mode. If the length of each segment is set so that $L>>1/\text{min}(|\beta_{k_s,k_i}-\beta_{k'_s,k'_i}|)$ then only the mode that is phase matched will be produced in a particular segment. Then, by varying the relative length of each segment, any linear combination of modes can be produced at the end of the array. This process could be used in every waveguide, engineering each to produce a different state $\psi_{n_p}$. Then driving the waveguides simultaneously can span the entire space defined by the  $\psi_{n_p}$. Since the $\psi_{n_p}$ can now, in the absence of degeneracies, be set to any state this means that the array can now be engineered to be all-optically reconfigured over any $N$ dimensional subspace.

So in conclusion waveguide arrays can be engineered to span any $N$ dimensional quantum output space. However this will generally require some optimization of the refractive index in each waveguide in order to remove degeneracies.

\section{Inferring the biphoton wavefunction from classical difference frequency generation measurements}

Fabrication of complex domain poling patterns in waveguide array devices will no doubt involve some systematic and random errors. Typically large numbers of devices with varied parameters will be fabricated on a single chip, and then the quantum properties of each device will be characterized. The task of characterising each device is very time consuming due to the difficulty of the quantum correlation measurements and the shear number of devices to be characterized. In order to efficiently test which devices operate correctly and which have unacceptably large fabrication errors one can use the 'stimulated emission tomography' \cite{Liscidini:2013-193602:PRL} approach based on classical difference frequency generation as an alternative to measurements of the quantum correlations caused by SPDC. This would provide a quick and efficient way of separating defective devices from correctly operating ones. Then further more detailed characterization can be preformed on devices that pass the initial classical DFG testing.

Here we show how to use classical difference frequency generation in a waveguide array to determine the quantum wavefunction that would be produced by SPDC. Spontaneous parametric down-conversion (SPDC) is the quantum analogue of difference frequency generation (DFG). The key difference being that DFG is stimulated by a specific photon state (the seed laser) where as SPDC is stimulated via quantum vacuum fluctuations. This implies it is possible to reconstruct the biphoton wavefunction that would be produced by SPDC by carefully choosing the seed field of DFG to mimic quantum vacuum fluctuations, as is demonstrated in \cite{Helt:2012-2199:JOSB, Liscidini:2013-193602:PRL}. We show how to apply this technique to arrays of coupled waveguides, allowing characterization of the SPDC wavefunction using classical DFG measurements.

First we derive the equation for DFG in an array of coupled waveguides. Then we show how seeding the array with a specific seed field profile allows reconstruction of an idler field proportional to the SPDC wavefunction. Finally we demonstrate that a complex seed profile is not actually needed to simulate SPDC. We can instead simply seed one waveguide at a time, then add a linear combination of the output idler fields together to achieve the same output that would be produced by the seed with complex spatial profile. This is due to the linearity of idler field with respect to the seed field in the case of negligible pump depletion.

\subsection{Difference frequency generation in a waveguide array}

The equation for the idler field produced by DFG in an array of coupled waveguides is
\begin{equation}
i\frac{\partial E_{n_{i}}}{\partial z} = -C \bigg[ E_{n_{i}+1} + E_{n_{i}-1} \bigg] \\
+ \sum\limits_{n_p = 1}^{N} i{E_{n_{s}}^{(s)}}^* A_{n_{p}}  d_{n_{p}}(z) e^{i \Delta \beta^{(0)} z} \delta_{n_{i},n_{p}} \delta_{n_{s},n_{p}}
\label{dif}
\end{equation}
Where we assume the fields of the pump ($A_{n_{p}}$), and seed ($E_{n_s}^{(s)}$) lasers are undepleted, and also that only one waveguide ($n_{p}$) is driven by the pump laser. We can solve this equation in the same way as for the SPDC case.
Using reciprocal space defined by
\begin{equation}
 E_{n_{i}} = \sum\limits_{k_{i}=1}^N \sin \left( \frac{\pi k_{i} n_{i}}{N+1} \right) f_{k_{i}},
\end{equation}
we get,


%

\begin{equation}
 f_{k_{i}}(L) =
e^{(i \beta_{k_{i}} L)}
 \int\limits_{0}^{L}dz e^{(i(\Delta \beta^{(0)} - \beta_{k_{i}}) z )} \sin \left( \frac{\pi k_{i} n_{p}}{N+1} \right) {E_{n_{p}}^{(s)}}^* A_{n_{p}}  d_{n_{p}}(z).
 \label{kspace_DFG}
\end{equation}
Here we have introduced $\beta_{k_{i}} = 2C\cos \left( \frac{\pi k_{i}}{N+1} \right) $, the contribution of the idler photons transverse modes to the phase matching conditions.

The seed field will also couple into neighbouring waveguides in the array, so the transverse profile of $E_{n_{s}}^{(s)}$ will evolve according to

\begin{equation}
i\frac{\partial E_{n_{s}}^{(s)}}{\partial z} = -C \bigg[ E_{n_{s}+1}^{(s)} + E_{n_{s}-1}^{(s)} \bigg].
\end{equation}
This is easily solved using the same method as was used to solve Eq. (\ref{dif}) giving

\begin{equation}
 f_{k_{s}}(z) =  f_{k_{s}}(0) \exp{ \left( 2i C \cos \left( \frac{\pi k_{s}}{N+1} \right) z \right)},
\end{equation}
in reciprocal space, and in position space

\begin{equation}
 E_{n_{s}}^{(s)}(z) =   \frac{2}{N+1}\sum\limits_{k_{s}=1}^N \sin \left(\frac{\pi k_{s} n_{s}}{N+1}\right)  f_{k_{s}}(0) \exp{\left(2i C \cos \left( \frac{\pi k_{s}}{N+1} \right) z\right)}.
 \label{seed_profile}
\end{equation}


Now we can substitute this profile for the seed field, $E_{n_{s}}^{(s)}(z)$, into the equation for the idler field,  $E_{n_{i}}(z)$, (Eq. ($\ref{kspace_DFG}$)). This gives,

\begin{equation}
 f_{k_{i}}^{(\text{DFG})}(L) = \frac{2}{N+1}e^{(i \beta_{k_{i}} L)} \sum\limits_{k_{s}=1}^N
 \int\limits_{0}^{L}dz
 e^{(i(\Delta \beta^{(0)} - \beta_{k_{i}} - \beta_{k_{s}}) z )}
 \sin \left( \frac{\pi k_{i} n_{p}}{N+1} \right)
 \sin \left( \frac{\pi k_{s} n_{p}}{N+1} \right)  f_{k_{s}}^{(s)}(0)^*  A_{n_{p}}  d_{n_{p}}(z),
\label{DFG_fsol}
\end{equation}
with $\beta_{k_{i}} = 2 C \cos \left( \frac{\pi k_{i}}{N+1} \right) $ and $\beta_{k_{s}} = 2 C \cos \left( \frac{\pi k_{s}}{N+1} \right) $. At this point it is interesting to compare the classical signal field $f_{k_{i}}$ to the wavefunction for the SPDC case,

\begin{equation}
f_{k_{s},k_{i}}^{(\text{SPDC})}(L)=e^{(i (\beta_{k_{i}}+\beta_{k_{s}})L)} \int\limits_{0}^{L} dz
 e^{(i( \Delta \beta^{(0)} - \beta_{k_{i}} - \beta_{k_{s}})z)}
   \sin \left( \frac{\pi k_{i} n_{p}}{N+1} \right)
    \sin \left( \frac{\pi k_{s} n_{p}}{N+1} \right) A_{n_{p}} d_{n_{p}}(z).
 \label{SPDC_f}
\end{equation}

Equations (\ref{SPDC_f}) and (\ref{DFG_f2}) are very similar, except in the classical DFG case all the seed modes, $k_{s}$, are summed over, eliminating the correlation between signal and idler. Now in Eq. ($\ref{DFG_f2}$) we can set $ f_{k_{s}}^{(s)}(0) = \delta_{k_{s},k_{s'}}$ so that the input seed is in some eigenstate of the coupling operator. This means that the output DFG state will be proportional to the SPDC state $f_{k_{s'},k_{i}}^{(\text{SPDC})}(L)$. So by making N measurements of the DFG output for $k_{s'} = 1,2,...,N$ we can construct the full SPDC wavefunction  $f_{k_{s},k_{i}}^{(\text{SPDC})}(L)$.
Also we give the initial seed mode a specific phase so that it matches the SPDC equation, hence the required seed profile is $ f_{k_{s}}^{(s)}(0) = f_{0}^{(s)} \delta_{k_{s},k_{s'}} e^{(-i\beta_{k_{s}} L)}$, this gives

\begin{equation}
 f_{k_{i}}^{(\text{DFG})}(L) = \frac{2}{N+1} e^{(i (\beta_{k_{i}}+\beta_{k_{s}}) L)}
 \int\limits_{0}^{L}dz
 e^{(i(\Delta \beta^{(0)} - \beta_{k_{i}} - \beta_{k_{s}}) z ) }
\sin \left( \frac{\pi k_{i} n_{p}}{N+1} \right)
 \sin \left( \frac{\pi k_{s} n_{p}}{N+1} \right)   A_{n_{p}} f_{0}^{(s)}  d_{n_{p}}(z),
\end{equation}
which is identical to equation (\ref{SPDC_f}) up to a constant factor, $ f_{0}^{(s)}$, reflecting the fact that DFG has a dependence on the intensity of the seed field, whereas SPDC does not.

\subsection{Reconstruction of $\Psi^{(\text{SPDC})}$ from $E^{(\text{DFG})}$}
\begin{figure}[]
\centering
\includegraphics[scale=0.6]{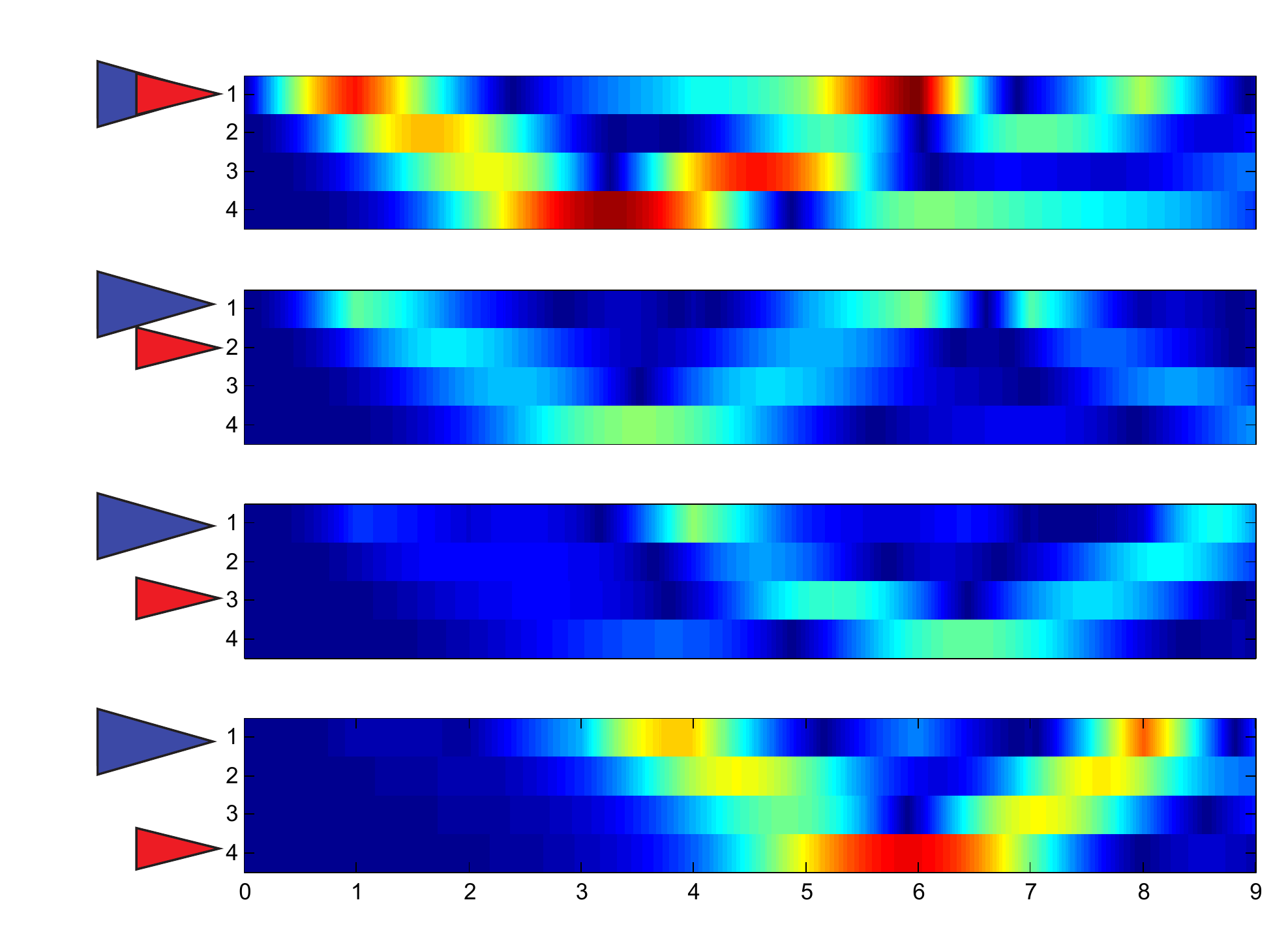}
\caption{Propagation of the idler field during difference frequency generation. The blue arrow shows the pumped waveguide, $n_p$, while the red arrow shows the seeded waveguide, $n_s'$.}
\label{fig:1}
\end{figure}

\begin{figure}[]
\centering
\begin{minipage}[t]{0.45\linewidth}
\includegraphics[scale=0.3]{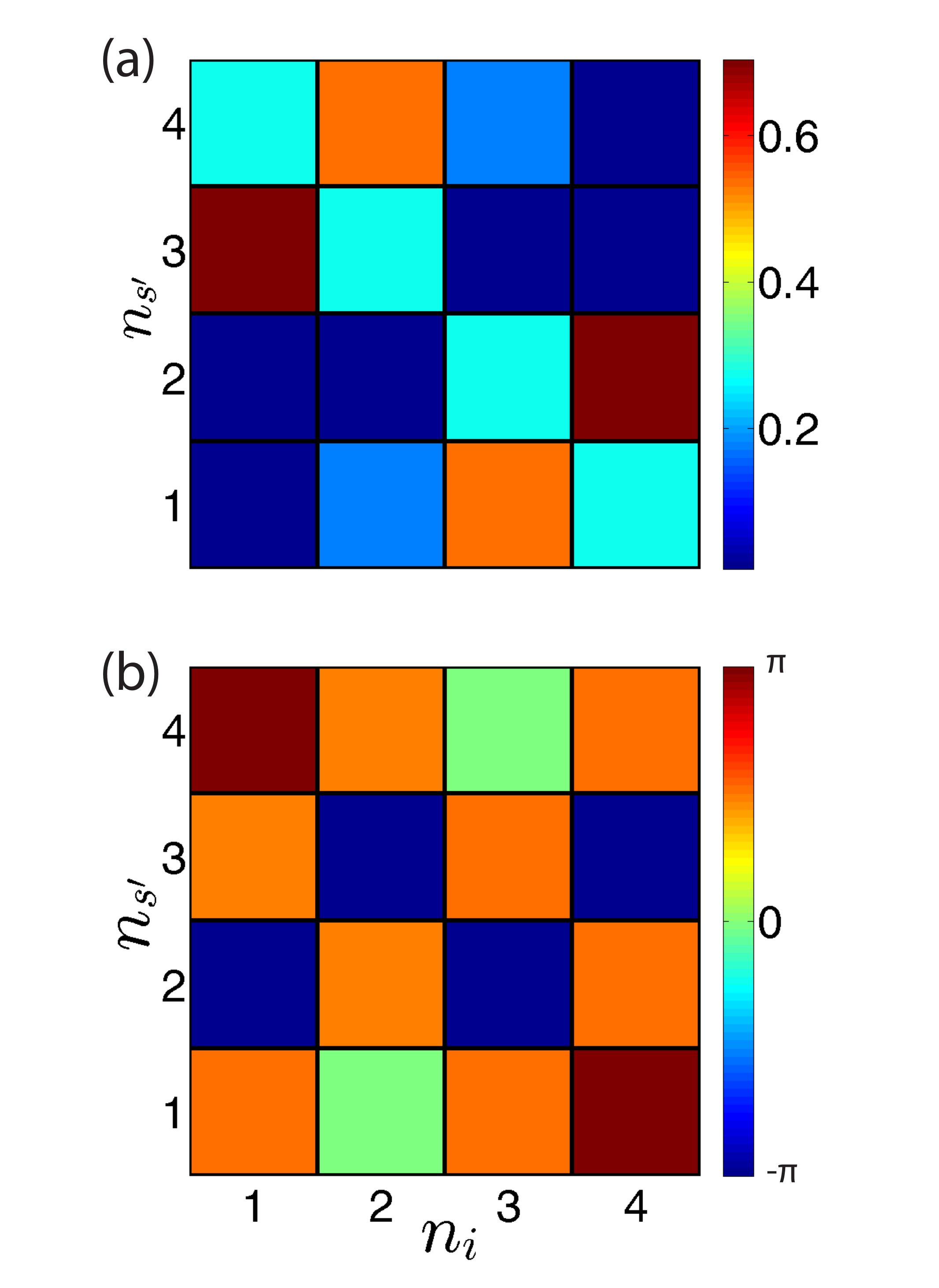}
\caption{ (a) A plot of $|E_{n_{i}n_{s'}}|^{2}$. The rows show the output idler field from figure (\ref{fig:1}). Each row corresponds the seed beam being coupled into a different waveguide, $n_{s'}$. (b) A plot of the phase of $E_{n_{i}n_{s'}}$.
  }
\label{f}
\end{minipage}
\quad
\begin{minipage}[t]{0.45\linewidth}
\includegraphics[scale=0.3]{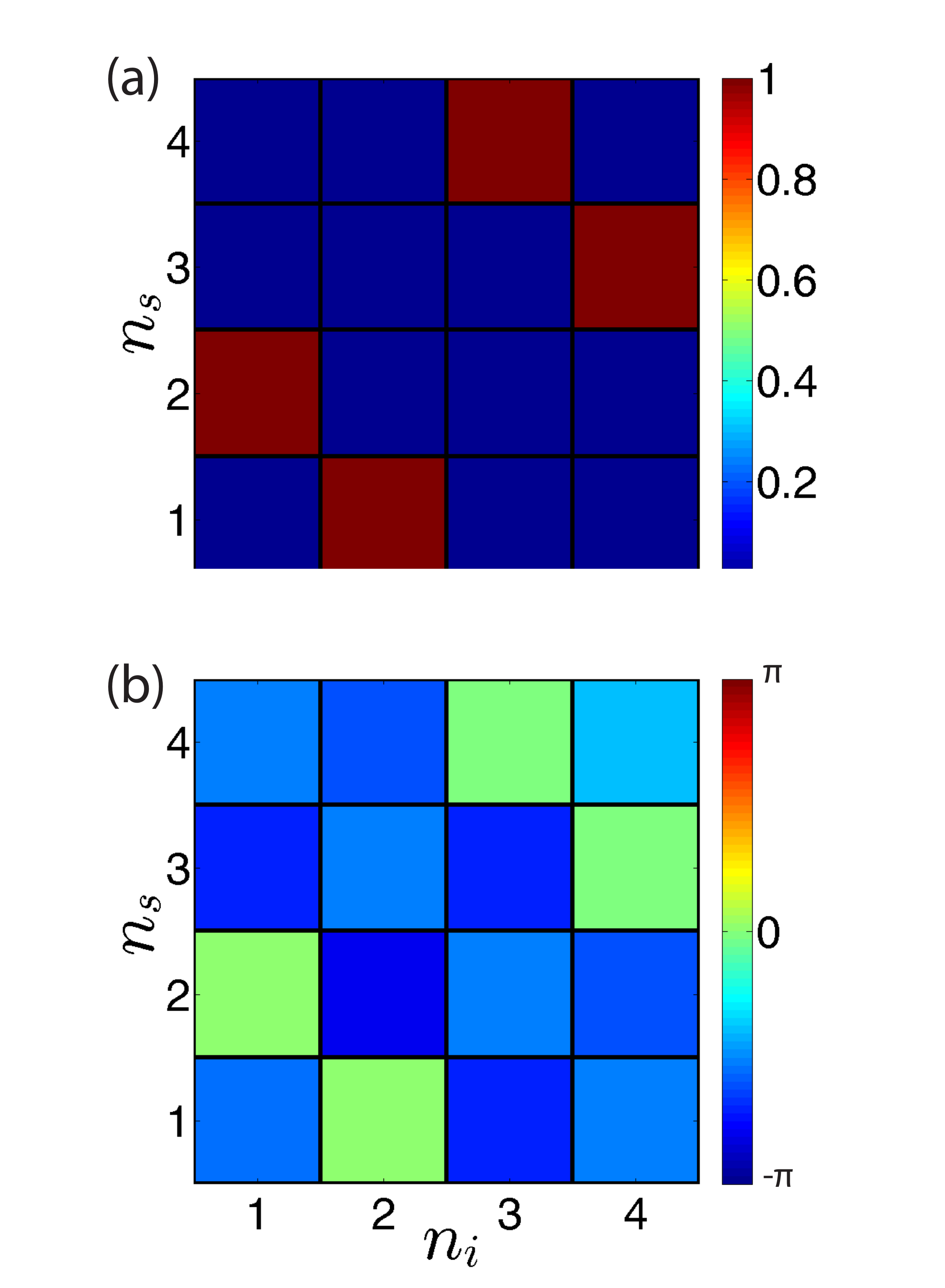}
\caption{A reconstruction of $\Psi{n_{i}n_{s}}$ using the transformation in equation (\ref{final_supp}) on the DFG output given in figure (\ref{f}). This confirms that the biphoton wavefunction produced by SPDC would be the desired Bell state, while avoiding the need to measure spacial entanglement between photons directly. (a) gives the intensity and (b) the phase of the reconstructed wavefunction, $\Psi{n_{i}n_{s}}$.}
\label{fig:minipage2}
\end{minipage}
\end{figure}

We have shown that DFG with the seed pulse in a particular transverse mode will be proportional to one row of the SPDC wavefunction (a phase factor is also required). So by taking multiple measurements of DFG ouput, with the seed in a different mode each time, it is possible to reconstruct the complete quantum mechanical wavefunction produced by SPDC. However in practise it would  be difficult to couple a seed laser into a waveguide array in a specific transverse mode.


 Typically the easiest quantity to measure would be the output idler field produced when seeding only one waveguide and pumping only one waveguide. So we will assume that this is the quantity that is actually measured, and show how to reconstruct the SPDC wavefunction from the measured values. We denote the measured outputs DFG outputs  $f_{k_{i},n_{s'}}^{(\text{meas.})}(L)$, where $k_{i}$ is the transverse mode of the output, and $n_{s'}$ the waveguide number that is seeded with the seed laser ($n_p$ denotes the pumped waveguide, but it is unnecessary to use this as an index).

 The expression for these measured values can be found by setting the initial seed profile in Eq. (\ref{seed_profile}) to $E_{n_s}^{(s)}(0) = \delta_{n_s,n_{s'}}$, then substituting into Eq. (\ref{DFG_fsol}) to get
 \begin{equation}
  f_{k_{i},n_{s'}}^{(\text{meas.})}(L) = \frac{2}{N+1}e^{(i \beta_{k_{i}} L)} \sum\limits_{k_{s}=1}^N
  \sin \left( \frac{\pi k_{i} n_{p}}{N+1} \right)
  \sin \left( \frac{\pi k_{s} n_{p}}{N+1} \right)   \sin \left( \frac{\pi k_s n_{s'}}{N+1} \right)
   \int\limits_{0}^{L}dz
    e^{(i(\Delta \beta^{(0)} - \beta_{k_{i}} - \beta_{k_{s}}) z )}
    A_{n_{p}}  d_{n_{p}}(z).
 \label{DFG_f2}
 \end{equation}

Note that setting $E_{n_s}^{(s)}(0) = \delta_{n_s,n_{s'}}$ implies that the seed magnitude and phase remain the same when changing the seed laser to different waveguides, $n_{s'}$, to measure different elements of $f_{k_{i},n_{s'}}^{(\text{meas.})}(L) $. In general it is not necessary to keep the intensity and phase the same, but any variations must be known to reconstruct the SPDC wavefunction. Here we assume for simplicity that  the phase and intensity of the seed remains constant regardless of which waveguide is seeded.

 Now once the idler output has been measured with the seed in each of the $N$ waveguides the SPDC wavefunction can be mathematically reconstructed from the measured values. This gives a reconstructed function $   f_{k_{i},k_{s'}}^{(\text{recon.})}(L)$ which is proportional to the SPDC wavefunction in Eq (\ref{SPDC_f}),
 \begin{equation}
 f_{k_{i},k_{s'}}^{(\text{recon.})}(L)= e^{(i \beta_{k_{s'}} L)}
 \sum_{n_{s'}=1}^N \sin \left( \frac{\pi n_{s'} k_{s'}}{N+1} \right) f_{k_{i}}^{(\text{meas.})}(L)
 \end{equation}

 Here we add together a superposition on the measured outputs , weighted by a $\sin$ function, and multiplied by a phase factor $e^{(i \beta_{k_{s'}} L)} $. The $k_{s'}$ argument determines which column of the SPDC wavefunction is reconstructed. To confirm that $f_{k_{i},k_{s'}}^{(\text{recon.})}(L)$ is a resonstruction of the SPDC wavefunction we substitute in the full equation for $f_{k_{i},n_{s'}}^{(\text{meas.})}(L)$ then simplify,
 \begin{multline}
 f_{k_{i},k_{s'}}^{(\text{recon.})}(L) =
 e^{(i \beta_{k_{s'}} L)}
  \sum_{n_{s'}=1}^N \sin \left( \frac{\pi n_{s'} k_{s'}}{N+1} \right)
  \frac{2}{N+1}e^{(i \beta_{k_{i}} L)} \sum\limits_{k_{s}=1}^N
   \sin \left( \frac{\pi k_{i} n_{p}}{N+1} \right)
   \sin \left( \frac{\pi k_{s} n_{p}}{N+1} \right)   \sin \left( \frac{\pi k_s n_{s'}}{N+1} \right)
   \\
   \times
    \int\limits_{0}^{L}dz
      e^{(i(\Delta \beta^{(0)} - \beta_{k_{i}} - \beta_{k_{s}}) z )}  A_{n_{p}}  d_{n_{p}}(z)
 \end{multline}
 summing over the seeded waveguide number, $n_{s'}$
 \begin{equation}
  f_{k_{i},k_{s'}}^{(\text{recon.})}(L) =
  e^{(i \beta_{k_{s'}} L)}
   \frac{2}{N+1}e^{(i \beta_{k_{i}} L)} \sum\limits_{k_{s'}=1}^N
    \sin \left( \frac{\pi k_{i} n_{p}}{N+1} \right)
    \sin \left( \frac{\pi k_{s'} n_{p}}{N+1} \right)
     \int\limits_{0}^{L}dz
       e^{(i(\Delta \beta^{(0)} - \beta_{k_{i}} - \beta_{k_{s'}}) z )}  A_{n_{p}}  d_{n_{p}}(z)
  \end{equation}
  This is proportional to the expression for the down converted wavefunction in Eq. (\ref{SPDC_f}). The same transformation can be applied to the real space version of the DFG output,

\begin{equation}
E_{n_{i}k_{s'}}^{(\text{recon.})}= e^{(i \beta_{k_{s'}} L)}
\sum_{n_{s'}=1}^N \sin \left( \frac{\pi n_{s'} k_{s'}}{N+1} \right)  E_{n_{i}n_{s'}}^{(\text{meas.})}
\end{equation}

Then the real space SPDC wavefunction can be recovered by transforming back from $k_{s'}$ space to $n_{s}$ space
\begin{equation}
E_{n_{i}n_{s'}}^{(\text{recon.})}=
\sum_{k_{s'}=1}^N \sin \left( \frac{\pi n_{s} k_{s'}}{N+1} \right)
 e^{(i \beta_{k_{s'}} L)}
\sum_{n_{s'}=1}^N \sin \left( \frac{\pi n_{s'} k_{s'}}{N+1} \right)
 E_{n_{i}n_{s'}}^{(\text{DFG})} \propto \Psi_{n_{i}n_{s'}}^{\text{(SPDC)}}
 \label{final_supp}
\end{equation}
This transformation allows reconstruction of the quantum SPDC wavefunction from measurements of the idler fields produced by DFG with only one waveguide seeded at a time. It would be straightforward to  generalize this procedure to the case of inhomogeneous refractive indices considered in section \ref{sec:removing}. So in principle fast classical characterization could be applied to arrays with varied refractive indices.



\subsection{Characterization procedure}

We have shown that a complete set of intensity and phase measurements of the DFG output of a device can be used to fully reconstruct the SPDC wavefunction. This will allow for quicker characterization of large numbers of devices. However this procedure will require measurements of phase difference between light in adjacent waveguides which will be much more difficult than intensity measurements. Hence a good first step for characterization would be to measure the intensity in each waveguide produced by DFG and compare this with the intensity predicted by Eq. (\ref{DFG_f2}) using the target fabrication parameters. Then, if the intensity measurements are in reasonable agreement with the target values, phase measurements could be taken to reconstruct the complete SPDC wavefunction. If the reconstructed wavefunction matches that target SPDC wavefunction then finally quantum correlation measurements of down converted photons can be used confirm the quantum properties of the device.

This three-tiered characterization procedure will allow the majority of defective devices to be quickly identified and discarded using classical measurement techniques. Hence less time will be spent preforming complex quantum correlation measurements on defective devices, considerably speeding up the characterization of quantum photonic chips.

\end{widetext}

\end{document}